\documentclass[
superscriptaddress,
nofootinbib,
twocolumn,
 amsmath,amssymb,
 aps,
pra,
notitlepage,
longbibliography
]{revtex4-1}

\usepackage{dcolumn}% Align table columns on decimal point
\usepackage{bm}% bold math
\usepackage{epsfig}
\usepackage{graphicx}
\usepackage{latexsym}
\usepackage{amsfonts}
\usepackage{setspace}
\usepackage{graphicx}
\usepackage{amsmath}
\usepackage{mathrsfs}
\usepackage{verbatim}
\usepackage{color}
\usepackage{SIunits}
\usepackage{hyperref}

\usepackage{physics}
\usepackage{soul}

\newcommand{\be}{\begin{equation}}
\newcommand{\ee}{\end{equation}}
\newcommand{\ba}{\begin{array}}
\newcommand{\ea}{\end{array}}
\newcommand{\bqa}{\begin{eqnarray}}
\newcommand{\eqa}{\end{eqnarray}}

\begin{document}

\title{Efficient bidirectional quantum frequency conversion between telecom and visible bands using programmable III-V nanophotonic waveguides}
% High-performance quantum frequency conversion using programmable unpoled nanophotonic waveguides

\author{Jierui Hu} 
\thanks{These authors contributed equally to this work.}
\affiliation{Holonyak Micro and Nanotechnology Laboratory and Department of Electrical and Computer Engineering, University of Illinois at Urbana-Champaign, Urbana, IL 61801 USA}
\affiliation{Illinois Quantum Information Science and Technology Center, University of Illinois at Urbana-Champaign, Urbana, IL 61801 USA}
\author{Hao Yuan} 
\thanks{These authors contributed equally to this work.}
\affiliation{Holonyak Micro and Nanotechnology Laboratory and Department of Electrical and Computer Engineering, University of Illinois at Urbana-Champaign, Urbana, IL 61801 USA}
\affiliation{Illinois Quantum Information Science and Technology Center, University of Illinois at Urbana-Champaign, Urbana, IL 61801 USA}
\author{Joshua Akin} 
\thanks{These authors contributed equally to this work.}
\affiliation{Holonyak Micro and Nanotechnology Laboratory and Department of Electrical and Computer Engineering, University of Illinois at Urbana-Champaign, Urbana, IL 61801 USA}
\affiliation{Illinois Quantum Information Science and Technology Center, University of Illinois at Urbana-Champaign, Urbana, IL 61801 USA}
\author{A. K. M. Naziul Haque} 
%\thanks{These authors contributed equally to this work.}
\affiliation{Holonyak Micro and Nanotechnology Laboratory and Department of Electrical and Computer Engineering, University of Illinois at Urbana-Champaign, Urbana, IL 61801 USA}
\affiliation{Illinois Quantum Information Science and Technology Center, University of Illinois at Urbana-Champaign, Urbana, IL 61801 USA}
\author{Yunlei Zhao} 
%\thanks{These authors contributed equally to this work.}
\affiliation{Holonyak Micro and Nanotechnology Laboratory and Department of Electrical and Computer Engineering, University of Illinois at Urbana-Champaign, Urbana, IL 61801 USA}
\affiliation{Illinois Quantum Information Science and Technology Center, University of Illinois at Urbana-Champaign, Urbana, IL 61801 USA}
\author{Kejie Fang} 
\email{kfang3@illinois.edu}
%\homepage{https://fang.ece.illinois.edu}
\affiliation{Holonyak Micro and Nanotechnology Laboratory and Department of Electrical and Computer Engineering, University of Illinois at Urbana-Champaign, Urbana, IL 61801 USA}
\affiliation{Illinois Quantum Information Science and Technology Center, University of Illinois at Urbana-Champaign, Urbana, IL 61801 USA}

\begin{abstract} 

Quantum frequency conversion (QFC) is essential for interfacing quantum systems operating at different wavelengths and for realizing scalable quantum networks. Despite extensive progress, achieving QFC with simultaneous high efficiency, low pump power, minimal noise, broad bandwidth, and pump-wavelength flexibility remains challenging. Here, we demonstrate efficient, low-noise, and bidirectional QFC between the telecom (1550-nm) and visible (780-nm) bands using unpoled indium gallium phosphide (InGaP) $\chi^{(2)}$ nanophotonic waveguides, eliminating the need for a long-wavelength pump. Leveraging the large nonlinear susceptibility of InGaP together with programmable modal-phase-matching control, we obtain record-low pump power (20 mW)---an order of magnitude lower than that in previous demonstrations using integrated thin-film waveguides---and record-high loss-inclusive normalized conversion efficiency among non-resonant QFC implementations. The platform maintains quantum coherence and entanglement of input photons with noise well below the single-photon level. 
These results mark a significant advance in integrated nonlinear photonics for high-performance QFC, facilitating the development of versatile and scalable quantum networks.

\end{abstract}
%\pacs{}

\maketitle

Future quantum networks will rely on interconnected quantum nodes composed of matter-based qubits and optical fiber links that transmit quantum information via single photons. Quantum frequency conversion (QFC) enables photons to shift between wavelengths---for example, bridging the telecom and visible bands---while preserving essential quantum properties such as coherence and entanglement \cite{kumar1990quantum, raymer2012manipulating, han2023quantum}. As a result, low-noise, high-efficiency QFC is indispensable for scalable quantum networks and hybrid quantum systems \cite{knaut2024entanglement, liu2024creation, stolk2024metropolitan}. QFC also facilitates single-photon detection in otherwise inaccessible spectral ranges, such as the infrared \cite{mancinelli2017mid,huang2021mid}, where direct detection is difficult or requires complex conditions. Beyond quantum technologies, frequency conversion finds wide-ranging applications in telecommunications, spectroscopy, and laser systems.

\begin{figure*}[!htb]
	\begin{center}
		\includegraphics[width=2\columnwidth]{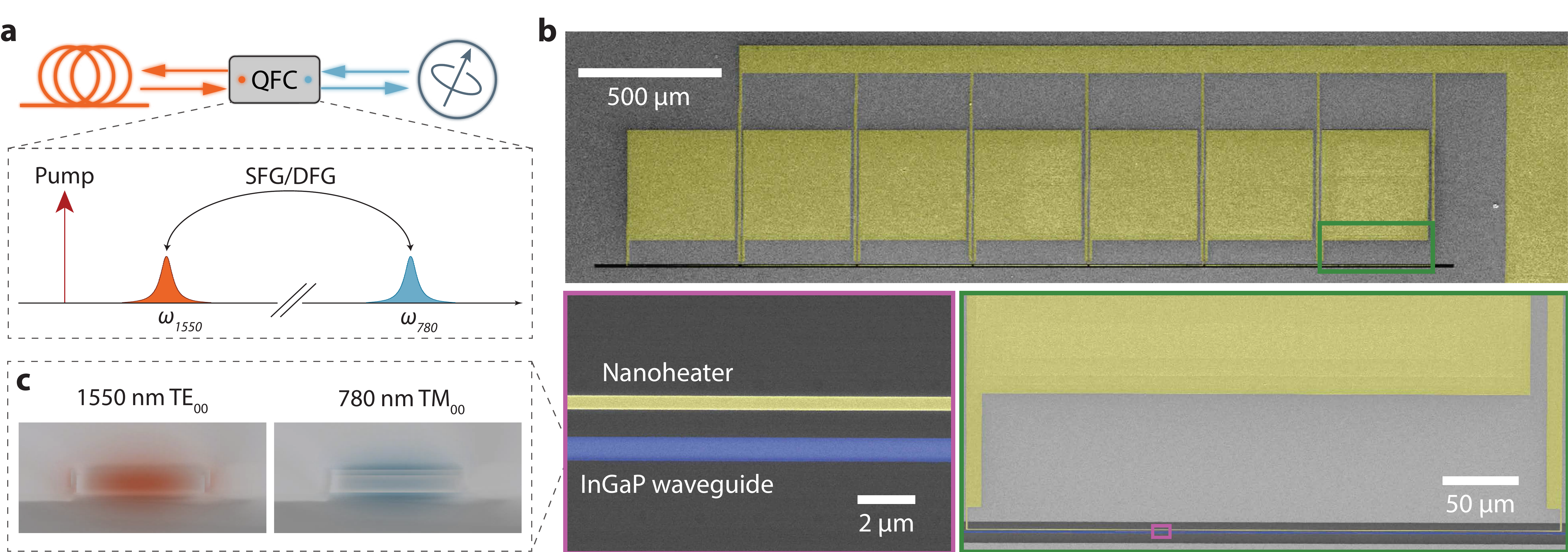}
		\caption{\textbf{Programmable InGaP nanophotonic waveguide for QFC.} 
			\textbf{a}. Schematic illustration of QFC between the 1550-nm and 780-nm bands with an ancillary pump via SFG or DFG.  \textbf{b}. False-colored scanning electron microscope (SEM) images of a phase-matching programmable InGaP nanophotonic waveguide integrated with a nanoheater array. Blue: InGaP waveguide. Yellow: gold electrode. \textbf{c}. Simulated mode profile overlaid on the SEM of the cross section of the waveguide.  }
		\label{fig::1}
	\end{center}
\end{figure*}

QFC is typically realized through nonlinear optical processes driven by ancillary pumps, including sum-frequency generation (SFG) \cite{rakher2010quantum, kaiser2019quantum, zheng2020integrated, niizeki2020two} and four-wave mixing \cite{li2016efficient, lu2019efficient,raghunathan2025telecom}, using $\chi^{(2)}$ or $\chi^{(3)}$ nonlinearities. While many demonstrations have relied on bulk nonlinear crystals \cite{ramelow2012polarization, kerdoncuff2021quantum, mann2023low,mann2023low,brevoord2025quantum} or optical fibers \cite{bonsma2022ultratunable}, integrated photonic platforms now offer a promising route to efficient QFC through chip-scale devices with strong mode confinement \cite{li2016efficient,singh2019quantum, wang2023quantum}. However, $\chi^{(3)}$ nonlinearities are inherently weak and demand high pump powers, which in turn generate excessive noise. Optical resonators can reduce pump requirements, but at the cost of narrow bandwidths and limited wavelength flexibility \cite{guo2016chip,lu2019efficient,singh2019quantum, wang2021efficient, chen2021photon,logan2023triply,raghunathan2025telecom}. QFC using $\chi^{(2)}$ nonlinearities has relied largely on periodically poled materials to realize quasi-phase matching, but poling disorder introduces statistically white noise from spontaneous parametric down-conversion (SPDC) of the pump \cite{pelc2010influence, fan2021photon, mann2024noise}. This necessitates long-wavelength pumps \cite{pelc2011long,fan2021photon} and often multistage conversion schemes \cite{saha2023low,yu2025efficient}, leading to degraded performance and increased system complexity. On the other hand, demonstrations of linear frequency conversion in unpoled $\chi^{(2)}$ materials have been rare and, to date, limited to the classical regime with modest efficiencies \cite{guo2016chip,wang2021efficient,logan2023triply,ahler2025second}. Thus, achieving QFC with simultaneous high efficiency, low pump power, minimal added noise, broad bandwidth, and pump-wavelength flexibility remains an open challenge.

Here, we demonstrate high-performance QFC between the telecom (1550-nm) and visible (780-nm) bands using unpoled indium gallium phosphide (InGaP) $\chi^{(2)}$ nanophotonic waveguides with programmable phase-matching control, simultaneously addressing multiple critical requirements for QFC. The III-V semiconductor InGaP platform \cite{zhao2022ingap, akin2024ingap, akin2024perspectives, thiel2024wafer} offers a large second-order nonlinear susceptibility ($\chi^{(2)} = 220$ pm/V), low optical loss, and a narrow Raman spectrum \cite{mediavilla2024composition}, while employing noninvasive modal phase matching. We leverage these features to reduce pump power demands, suppress parasitic noise, and relax constraints on pump-wavelength configuration of QFC. We further implement programmable phase-matching control to compensate thin-film nonuniformities and optimize both SFG and difference-frequency generation (DFG) for bidirectional conversion. Our device achieves record-low pump power and record-high loss-inclusive normalized conversion efficiency among non-resonant QFC implementations. Importantly, on-chip noise remains well below the single-photon level---even given the spectral proximity of the pump and signal---preserving the coherence and entanglement of input photons in our experiment. By combining efficiency, low power, bidirectionality, and noise suppression in a scalable integrated photonic platform, our approach addresses a critical need for advancing the development of versatile and hybrid quantum networks.

\noindent\textbf{Unpoled $\chi^{(2)}$ waveguide with programmable phase matching}\\
The InGaP nanophotonic waveguide is designed to support the modal-phase-matched 1550-nm-band transverse electric (TE$_{00}$) mode and 780-nm-band transverse magnetic (TM$_{00}$) mode (Fig.~\ref{fig::1}). QFC between these modes occurs via SFG or DFG with an ancillary 1550-nm-band pump, satisfying the frequency and phase-matching conditions $\omega_{1550} + \omega_p = \omega_{780}$ and $k_{1550} + k_p = k_{780}$, where $\omega_{1550}$ and $k_{1550}$ denote the angular frequency and wavevector of the 1550-nm-band light, respectively, and similarly for the 780-nm-band light and pump. In the weak-pump limit, the simulated normalized SFG efficiency is $\eta_{\mathrm{SFG}}\equiv\frac{P_{780}}{P_{1550}P_pL^2} = 520,000\%$/W/cm$^2$ for a lossless, perfectly phase-matched InGaP nanophotonic waveguide. 

However, one major challenge in utilizing integrated photonic waveguides for $\chi^{(2)}$ nonlinear processes lies in the thickness nonuniformity of the thin-film material \cite{chen2024adapted, li2024advancing}, which results in accumulated phase mismatch along the waveguide length and thus degraded conversion efficiency. To overcome this, we developed an in situ phase-matching tuning technique, using a nanoheater array integrated adjacent to the waveguide (Fig.~\ref{fig::1}b) (see the Supplementary Information (SI) for device fabrication). By precisely controlling the temperature of individual waveguide segments with a voltage of only up to a few volts, we can counteract the effect of thickness variations and correct the phase-matching condition along the full waveguide length (SI). The nanoheater array is programmable through the applied voltage to optimize the phase-matching conditions for different nonlinear processes and different wavelengths.

For a phase-matched waveguide, the photon-flux-based conversion efficiency of an SFG or DFG process is given by (SI)
\be\label{CE}
\eta=\eta_{\mathrm{max}}\sin^2\left(\frac{1}{4}\sqrt{8\eta_{\mathrm{SFG}}P_p-(\beta-\alpha)^2}L\right),
\ee
where 
\be
\eta_{\mathrm{max}}=\frac{8\eta_{\mathrm{SFG}}P_pe^{-\frac{(\alpha+\beta)L}{2}}}{8\eta_{\mathrm{SFG}}P_p-(\beta-\alpha)^2},
\ee 
$L$ is the waveguide length, $P_p$ is the pump power, and $\alpha$ and $\beta$ are the optical losses for the 1550-nm and 780-nm bands, respectively. This model assumes negligible pump loss. From these expressions, the peak conversion efficiency scales as $e^{-(\alpha+\beta)L/2}$, i.e., the geometric average of the  transmission efficiency of the two modes, whereas the pump power required to reach it scales as $1/L^2$ and is inversely proportional to $\eta_{\mathrm{SFG}}$.

\begin{figure*}[!htb]
	\begin{center}
		\includegraphics[width=2\columnwidth]{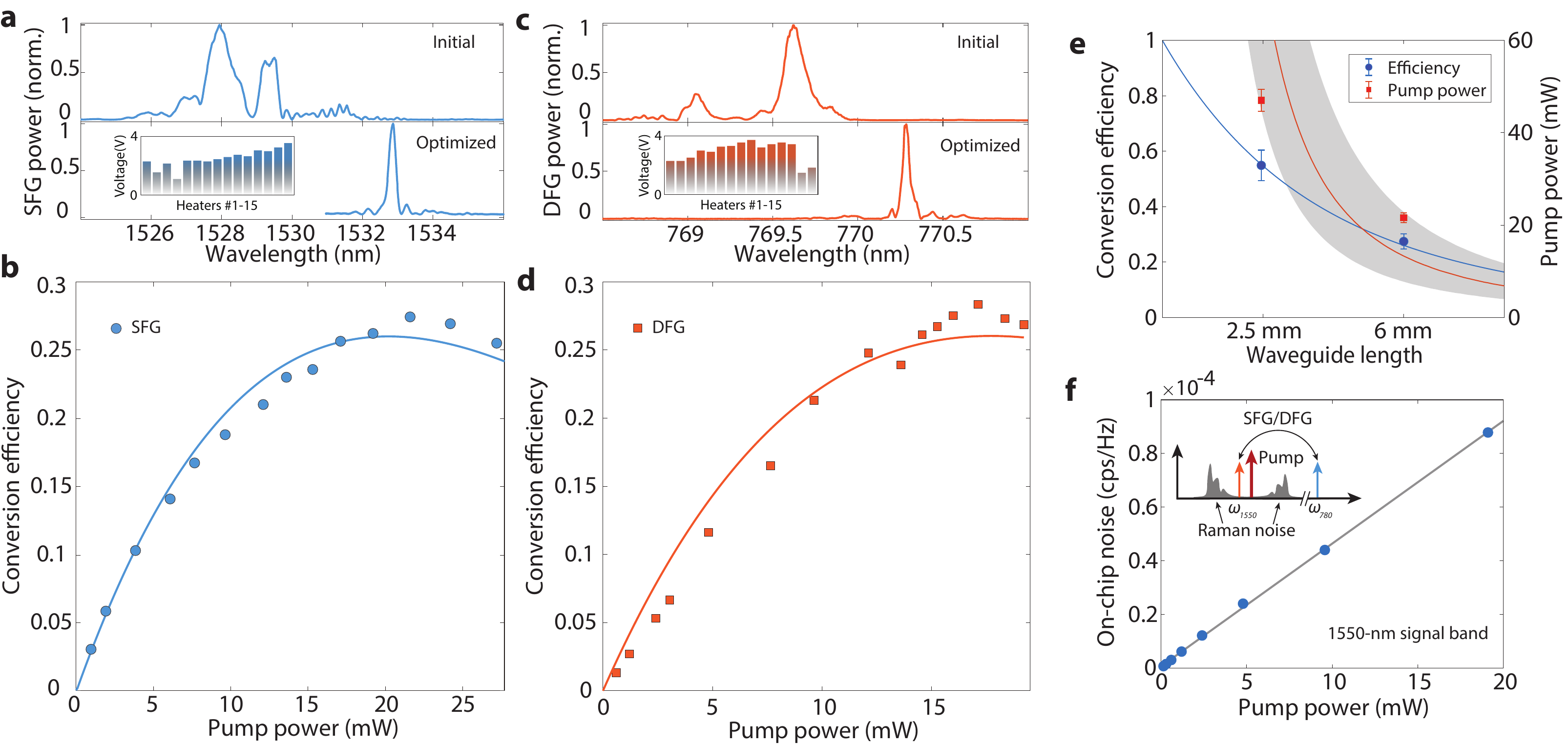}
		\caption{\textbf{Phase-matching optimized SFG and DFG.} 
			\textbf{a}. Initial and optimized SFG spectra of a 6-mm waveguide. Inset shows the applied voltages of nanoheaters to optimize the phase matching. \textbf{b}. Internal SFG conversion efficiency versus the pump power after optimization. Solid line is model fitting.  \textbf{c}. Initial and optimized DFG spectra. \textbf{d}. Internal DFG conversion efficiency versus the pump power after optimization.  \textbf{e}. Peak conversion efficiency and corresponding pump power of the 2.5-mm and 6-mm waveguides. Error bars are due to uncertainty in the fiber-optic coupling efficiency. Solid lines are theoretical modeling. \textbf{f}. On-chip noise flux in the 1550-nm band at the wavelength separated from the pump by 20 nm ($-2.5$ THz) versus the pump power of the 6-mm waveguide. The line is a linear fitting. Inset illustrates the pump-induced Raman noise and QFC wavelengths.}
		\label{fig::2}
	\end{center}
\end{figure*}

\begin{figure*}[!htb]
	\begin{center}
		\includegraphics[width=2\columnwidth]{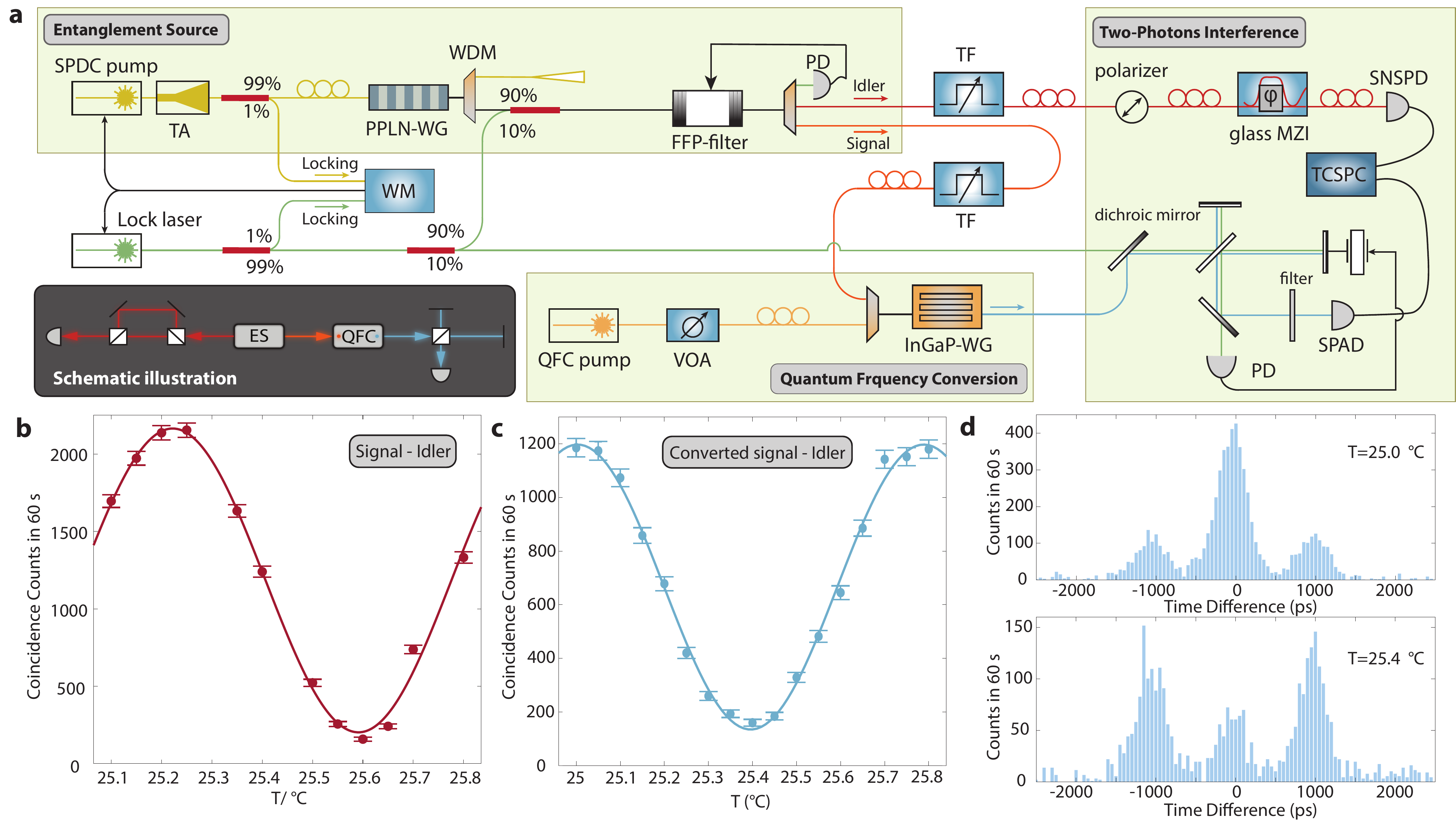}
		\caption{\textbf{Time--energy entanglement of telecom and visible photons via QFC.} 
			\textbf{a}. Experimental schematic and setup. ES: entanglement source. TA: tapered amplifier. PPLN-WG: periodically-poled lithium niobate waveguide. WDM: wavelength-division multiplexer. FFP filter: fiber Fabry-Perot filter. PD: photodetector. TF: tunable filter. WM: wavelength meter. VOA: variable optical attenuator. SNSPD: superconducting nanowire single-photon detector. TCSPC: time-correlated single-photon counting module. SPAD: single-photon avalanche diode detector. \textbf{b}. Coincidence counts of the signal and idler in 150-ps binwidth integrated in 60 s for various glass MZI temperatures. \textbf{c}. Coincidence counts of the converted signal and idler. \textbf{d}. Time--difference histograms of the converted signal and idler for two glass MZI temperatures. Time-bin width is 50 ps. }
		\label{fig::3}
	\end{center}
\end{figure*}

\noindent\textbf{Optimized SFG and DFG}\\
Figure~\ref{fig::2}a shows the SFG spectra of a 6-mm-long waveguide under a 1550-nm pump before and after phase-matching optimization, resulting in a near-sinc$^2$ SFG transfer function at 1533 nm with a bandwidth of 31 GHz. The tuning voltages of the array of 15 nanoheaters for the optimization are shown in the inset. The measured SFG conversion efficiency versus pump power after phase-matching optimization is displayed in Fig.~\ref{fig::2}b. We use adiabatically tapered fiber couplers with efficiencies of 80\% and 50\% for the 1550-nm TE$_{00}$ and 780-nm TM$_{00}$ modes, respectively.
The measured peak internal conversion efficiency is nearly 30\%, limited by the loss of the 780-nm TM$_{00}$ mode of 20 dB/cm (SI), whereas the loss of the 1550-nm TE$_{00}$ mode is approximately 1 dB/cm. This is consistent with the measured quality factor of 780-nm-band microring resonances ($Q_{780} \sim 10^5$) \cite{akin2024perspectives} and a simulated group index of $n_g = 5.7$ for the 780-nm TM$_{00}$ mode (i.e., $\beta = n_g \omega / c Q$). The experimentally inferred $\eta_{\mathrm{SFG}}=70,000\%$/W/cm$^2$, by fitting the conversion efficiency curve via Eq. \ref{CE}, is lower than the theoretical value, leading to a higher pump power ($\sim20$ mW) to reach the peak efficiency. This discrepancy arises from the residual phase mismatch at length scales shorter than those of individual nanoheaters and can be improved with more discrete nanoheaters (SI). 

Owing to the different waveguide losses for the 1550-nm and 780-nm modes and residual phase mismatch, the conversion efficiency of DFG with the same pump direction as the SFG is different from that of the SFG process and must be separately optimized. In contrast, the conversion efficiencies of DFG and SFG with opposite pump directions are the same due to time reversal (see extra data in the SI). Fig.~\ref{fig::2}c shows the DFG spectra under 1532-nm pump before and after its phase-matching optimization for the same 6-mm waveguide and same pump direction as SFG, and Fig.~\ref{fig::2}d shows the measured DFG conversion efficiency versus pump power after optimization. The peak efficiency and the corresponding pump power are both similar to those of the independently optimized SFG process.  

To explore scaling behavior, we measured devices of different lengths. A 2.5-mm-long phase-matching-optimized waveguide exhibits approximately double the peak conversion efficiency of the 6-mm device to 55\%, which is consistent with the exponential scaling with the waveguide loss (Fig.~\ref{fig::2}e), and a bandwidth of 50 GHz. The pump power (50 mW) required for peak efficiency does not follow the $1/L^2$ trend though, due to less phase mismatch accumulation in the short 2.5-mm device with inferred $\eta_{\mathrm{SFG}}=175,000\%$/W/cm$^2$ (SI). 
We also measured the on-chip noise in the 1550-nm band of the 6-mm waveguide without signal injection (SI), which is linear in pump power and reaches approximately $10^{-4}$ counts per second per hertz (cps/Hz) at a 20-mW pump power and at the wavelength separated from the pump by $\pm20$ nm ($\mp2.5$ THz) (Fig.~\ref{fig::2}f). The noise is much less than the single-photon limit of $\sim 1$ cps/Hz. The corresponding up-converted noise flux is approximately $2\times10^{-5}$ cps/Hz according to the calculation (SI). The linear noise is attributed to spontaneous Raman scattering of the pump in the waveguide (see the SI for the modeling). We also observed that, at wavelengths within approximately 10 nm of the pump, the noise scales quadratically with the pump power, which originates from the degenerate spontaneous four-wave mixing of the pump (SI). In contrast to periodically poled materials, the unpoled III-V platform is free of broadband SPDC noise. 
Moreover, the QFC wavelength can be tuned by varying the pump wavelength or device temperature \cite{akin2024ingap}, with a simulated SFG phase-matching wavelength range of $>1000$ nm for the fundamental mode and $>30$ nm for the SFG mode for the fabricated waveguide (SI). 

\begin{figure*}[!htb]
	\begin{center}
		\includegraphics[width=2\columnwidth]{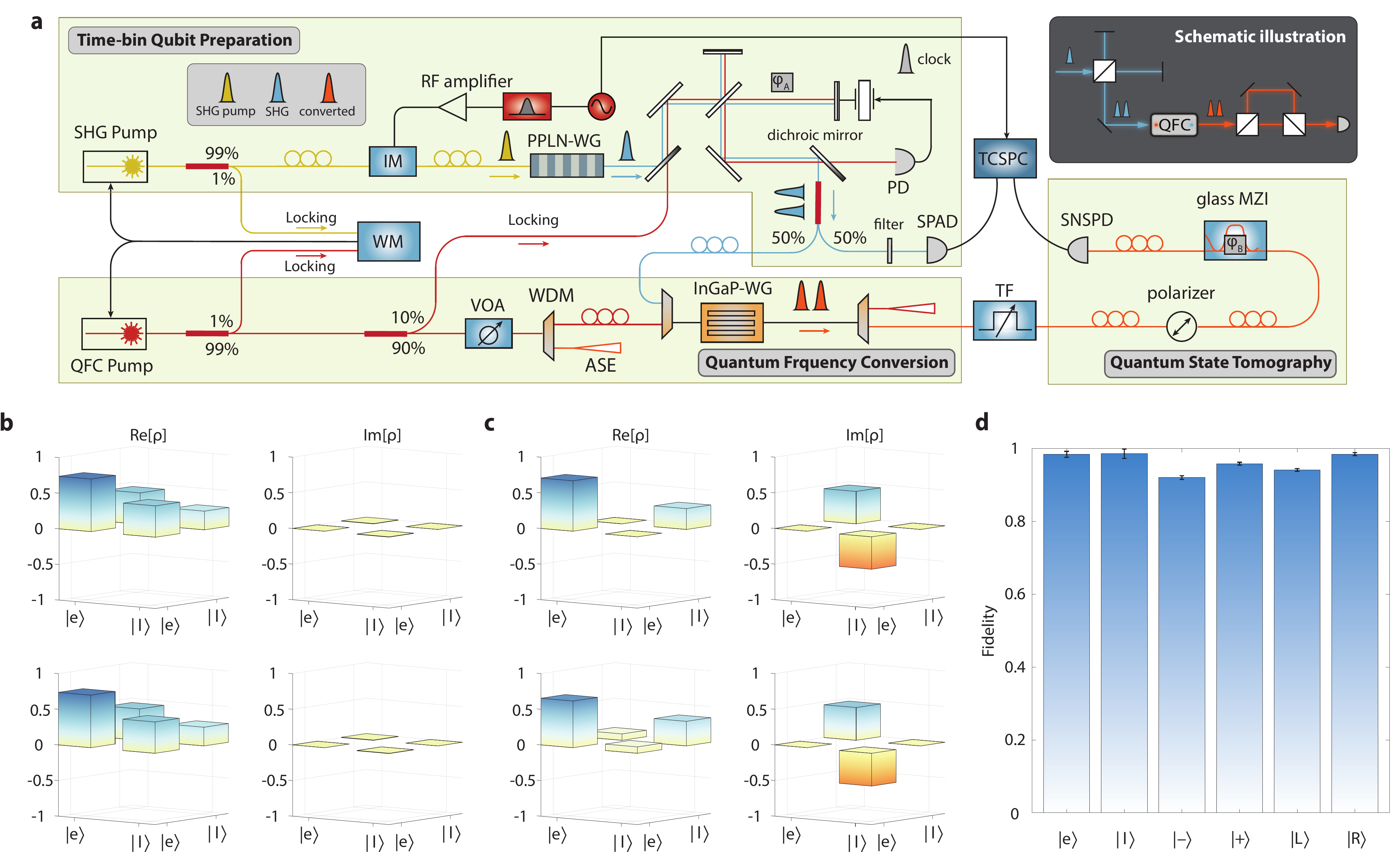}
		\caption{\textbf{Down-conversion of time-bin qubit.} 
			\textbf{a}. Experimental schematic and setup. IM: intensity modulator. ASE: amplified spontaneous emission. For other acronyms see Fig. \ref{fig::3}. \textbf{b}.  Density matrices of $\left|+\right\rangle$-like state before (top) and after (bottom) QFC.  \textbf{c}. Density matrices of $\left|R\right\rangle$-like state before (top) and after (bottom) QFC. \textbf{d}. Fidelities of the six basis states after QFC.}
		\label{fig::4}
	\end{center}
\end{figure*}

\noindent\textbf{Single-photon up-conversion}\\
We demonstrate that bidirectional QFC in InGaP nanophotonic waveguides preserves both quantum entanglement and single-photon coherence. First, we verify that the time--energy entanglement of photon pairs generated via SPDC is preserved after one photon is up-converted. The experimental setup is depicted in Fig.~\ref{fig::3}a. A commercial periodically poled LiNbO$_3$ waveguide, pumped by a 780-nm-band continuous-wave laser, generates time--energy entangled photons in the telecommunications band. Non-degenerate signal and idler photons are spectrally selected via a tunable fiber Fabry--Perot filter with a 2.3-GHz bandwidth. The signal photon is coupled into the InGaP waveguide and up-converted to the 780-nm band via SFG. We perform time-resolved two-photon interference measurements to verify the time--energy entanglement. We first confirm the entanglement of the unconverted SPDC photons via a pair of glass Mach--Zehnder interferometers (MZIs) with a path delay of 1~ns, which is much longer than the coherence time of the SPDC photons. The phase of the MZI is controlled by temperature. Fig.~\ref{fig::3}b shows the two-photon interference fringe obtained by varying the temperature of one MZI, yielding a visibility of 83\%, above the Clauser--Horne limit of $1/\sqrt{2}$ \cite{clauser1974experimental}. After up-conversion, the two-photon interference is measured via one MZI for the idler and one free-space Michelson interferometer for the up-converted signal. Fig.~\ref{fig::3}c shows the resulting fringe, and time--difference histograms for two interferometer phases are plotted in Fig.~\ref{fig::3}d. The visibility remains at 80\%, confirming that entanglement is preserved after QFC. The two-photon fringe visibility before and after QFC are both limited primarily by the interferometer visibility (SI).

\noindent\textbf{Single-photon down-conversion}\\
In the down-conversion direction, we verify the preservation of single-photon coherence for a time-bin qubit. The experimental setup is depicted in Fig.~\ref{fig::4}a. A 780-nm-band time-bin qubit with a temporal delay $\tau = 1$ ns is created by second-harmonic generation of an attenuated 1550-nm laser pulse, followed by passing through a phase-tunable unbalanced Michelson interferometer. The qubit is converted to the 1550-nm band via DFG in the InGaP waveguide. After the residual pump is filtered out, quantum state tomography is performed on the down-converted qubit (SI). Figs.~\ref{fig::4}b and c show the reconstructed density matrices for the $\left|+\right\rangle$- and $\left|R\right\rangle$-like states, before and after conversion. Fig.~\ref{fig::4}d summarizes the fidelity results for six basis states---$\left|e\right\rangle$, $\left|l\right\rangle$, $\left|+\right\rangle$, $\left|-\right\rangle$, $\left|L\right\rangle$, and $\left|R\right\rangle$---with an average fidelity of $\mathcal{F} = 96.2 \pm 0.6\%$. The dominant contribution to infidelity arises from limited interferometer visibility and phase errors, with uncertainty contributed from phase fluctuations and photon shot noise (SI).

\begin{figure}[!htb]
	\begin{center}
		\includegraphics[width=1\columnwidth]{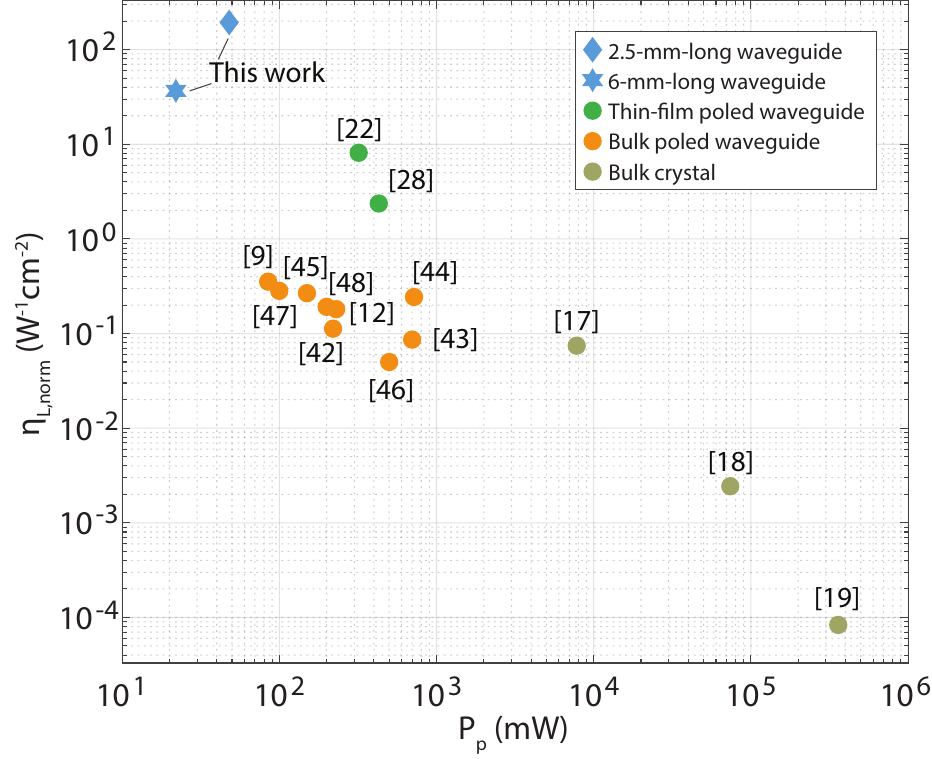}
		\caption{\textbf{Comparison of non-resonant $\chi^{(2)}$-based QFC performance.} Loss-inclusive normalized conversion efficiency is defined as $\eta_\mathrm{L,norm}\equiv\eta_{\mathrm{max}}/P_p/L^2$, where $\eta_{\mathrm{max}}$ is the peak internal  conversion efficiency, $P_p$ is the corresponding pump power, and $L$ is the device length. Thin-film poled waveguide:  \cite{fan2021photon,wang2023quantum}. Bulk poled waveguide:  \cite{morrison2021bright,niizeki2020two, bock2018high, rakher2010quantum,  ates2012two,zaske2012visible,vonChamier2025low,dreau2018quantum,ikuta2018polarization}.  Bulk crystal:  \cite{mann2023low,kerdoncuff2021quantum,brevoord2025quantum}. } 
		\label{fig::5}
	\end{center}
\end{figure}

In summary, we have demonstrated the first high-performance QFC that simultaneously meets all critical requirements using unpoled nanophotonic waveguides. The combination of strong second-order nonlinearity, noninvasive modal phase matching, and programmable phase-matching control enables high internal conversion efficiency with record-low pump power and minimal added noise. To benchmark the QFC performance, we introduce the loss-inclusive normalized conversion efficiency, $\eta_\mathrm{L,norm} \equiv \eta_{\mathrm{max}}/P_p/L^2$, where $\eta_{\mathrm{max}}$ is the peak internal conversion efficiency, $P_p$ is the corresponding pump power, and $L$ is the device length. While analogous to $\eta_\mathrm{SFG}$, this metric incorporates optical loss and is defined in the strong-pump regime. The $\eta_\mathrm{L,norm}$ of our work exceeds that of all non-resonant $\chi^{(2)}$-based QFC reported to date (see Fig. \ref{fig::5}). The III--V semiconductor platform also offers the potential for directly integrating pump lasers with the QFC component toward a turn-key system \cite{majid2015first}. Furthermore, the cutoff wavelength of InGaP ($\sim$650 nm) is compatible with a wide range of quantum emitters \cite{akin2024perspectives}, including quantum dots, color centers, and neutral atoms, making this platform promising for facilitating scalable and versatile quantum networks and hybrid quantum systems.

\vspace{2mm}
\noindent\textbf{Acknowledgements}\\ 
This work is supported by US National Science Foundation under Grant No. 2223192 and QLCI-HQAN (Grant No. 2016136) and U.S. Department of Energy Office of Science National Quantum Information Science Research Centers.

\noindent\textbf{Author contributions}\\ 
A.K.M.N.H., J.H., Y.Z., J.A. designed the device.  J.A., J.H. fabricated the device. J.H., H.Y, J.A. performed the experiment and analyzed the data. K.F. supervised the project. J.H., H.Y., J.A., K.F. wrote the paper.

%\noindent\textbf{Additional information}\\
%Supplementary information is available in the online version of the paper.  Reprints and permissions are available at www.nature.com/reprints.  The authors declare no competing financial interests.  Correspondence and requests for materials should be sent to KF (kfang3@illinois.edu). \\
%
%\noindent\textbf{Competing financial interests}\\
%The authors declare no competing financial interests.
 
\end{document}

\title{Supplementary Information for: High-performance quantum frequency conversion using programmable unpoled nanophotonic waveguides}

\author{Jierui Hu} 
\thanks{These authors contributed equally to this work.}
\affiliation{Holonyak Micro and Nanotechnology Laboratory and Department of Electrical and Computer Engineering, University of Illinois at Urbana-Champaign, Urbana, IL 61801 USA}
\affiliation{Illinois Quantum Information Science and Technology Center, University of Illinois at Urbana-Champaign, Urbana, IL 61801 USA}
\author{Hao Yuan} 
\thanks{These authors contributed equally to this work.}
\affiliation{Holonyak Micro and Nanotechnology Laboratory and Department of Electrical and Computer Engineering, University of Illinois at Urbana-Champaign, Urbana, IL 61801 USA}
\affiliation{Illinois Quantum Information Science and Technology Center, University of Illinois at Urbana-Champaign, Urbana, IL 61801 USA}
\author{Joshua Akin} 
\thanks{These authors contributed equally to this work.}
\affiliation{Holonyak Micro and Nanotechnology Laboratory and Department of Electrical and Computer Engineering, University of Illinois at Urbana-Champaign, Urbana, IL 61801 USA}
\affiliation{Illinois Quantum Information Science and Technology Center, University of Illinois at Urbana-Champaign, Urbana, IL 61801 USA}
\author{A. K. M. Naziul Haque} 
%\thanks{These authors contributed equally to this work.}
\affiliation{Holonyak Micro and Nanotechnology Laboratory and Department of Electrical and Computer Engineering, University of Illinois at Urbana-Champaign, Urbana, IL 61801 USA}
\affiliation{Illinois Quantum Information Science and Technology Center, University of Illinois at Urbana-Champaign, Urbana, IL 61801 USA}
\author{Yunlei Zhao} 
%\thanks{These authors contributed equally to this work.}
\affiliation{Holonyak Micro and Nanotechnology Laboratory and Department of Electrical and Computer Engineering, University of Illinois at Urbana-Champaign, Urbana, IL 61801 USA}
\affiliation{Illinois Quantum Information Science and Technology Center, University of Illinois at Urbana-Champaign, Urbana, IL 61801 USA}
\author{Kejie Fang} 
\email{kfang3@illinois.edu}
%\homepage{https://fang.ece.illinois.edu}
\affiliation{Holonyak Micro and Nanotechnology Laboratory and Department of Electrical and Computer Engineering, University of Illinois at Urbana-Champaign, Urbana, IL 61801 USA}
\affiliation{Illinois Quantum Information Science and Technology Center, University of Illinois at Urbana-Champaign, Urbana, IL 61801 USA}

\maketitle

\tableofcontents

\newpage

\section{Device fabrication}

\begin{figure*}[!htb]
	\begin{center}
		\includegraphics[width=1\columnwidth]{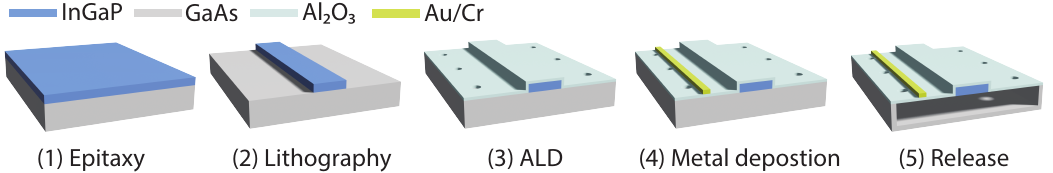}
		\caption{Illustration of the device fabrication process. }
		\label{fig::S0}
	\end{center}
\end{figure*}

The device fabrication process is illustrated in Fig. \ref{fig::S0}. The devices are fabricated from the 112 nm thick disordered In$_{0.5}$Ga$_{0.5}$P thin film grown on GaAs substrate by metal-organic chemical vapor deposition (T = 545 C, V/III = 48, precursors: trimethylindium, trimethylgallium and PH$_3$). The device pattern is defined using electron beam lithography and 150 nm thick negative tone resist hydrogen silsesquioxane (HSQ). A 20 nm thick layer of silicon dioxide is deposited on InGaP via plasma-enhanced chemical vapor deposition (PECVD) to promote the adhesion of HSQ. The device pattern is transferred to InGaP layer via inductively coupled plasma reactive-ion etch (ICP-RIE) using a mixture of Cl$_2$/CH$_4$/Ar gas. After a short buffered oxide etch to remove the residual oxide (both HSQ and PECVD oxide), a layer of 35 nm thick aluminum oxide is deposited on the chip via atomic layer deposition. A second electron beam lithography and subsequent ICP-RIE using CHF$_3$ gas are applied to pattern etch-through holes in the aluminum oxide layer for the undercut of the InGaP device. Next, a third electron beam lithography followed by electron-beam evaporation of 5 nm thick chromium and 20 nm thick gold is performed to define the electrodes. Finally, the InGaP device is released from the GaAs substrate using citric acid-based selective etching.

\section{Coupled-mode equation for frequency conversion}
Consider a $\chi^{(2)}$ waveguide supporting three optical modes, $a$, $b$, and $c$, satisfying $\omega_a + \omega_b = \omega_c$. We denote the phase mismatch $\Delta k=k_c-k_a - k_b$. The coupled-mode equations for the parametric frequency conversion (with mode $a$ being pumped) are given by
\begin{align}
\frac{\partial b}{\partial z}=-igc e^{-i\Delta k z}-\frac{\alpha}{2}b,\\
\frac{\partial c}{\partial z}=-igb e^{i\Delta k z}-\frac{\beta}{2}c,
\end{align}
where $g=\sqrt{\eta_{\mathrm{SFG}}P_p/2}$ for near-degenerate $a$ and $b$ modes, $\alpha$($\beta$) is the loss of mode $b(c)$, and $|b|^2(|c|^2$) is the photon flux of mode $b$($c$). The coupled-mode equation can be solved analytically. When $\Delta k=0$ and there is input in only one mode, the conversion efficiency in terms of photon flux is found to be
\be\label{CE}
\eta=\frac{16g^2e^{-\frac{(\alpha+\beta)L}{2}}}{16g^2-(\beta-\alpha)^2}\sin^2\left(\frac{1}{4}\sqrt{16g^2-(\beta-\alpha)^2}L\right),
\ee
where $L$ is the waveguide length.

\section{QFC phase-matching wavelength}

\begin{figure*}[!htb]
	\begin{center}
		\includegraphics[width=0.5\columnwidth]{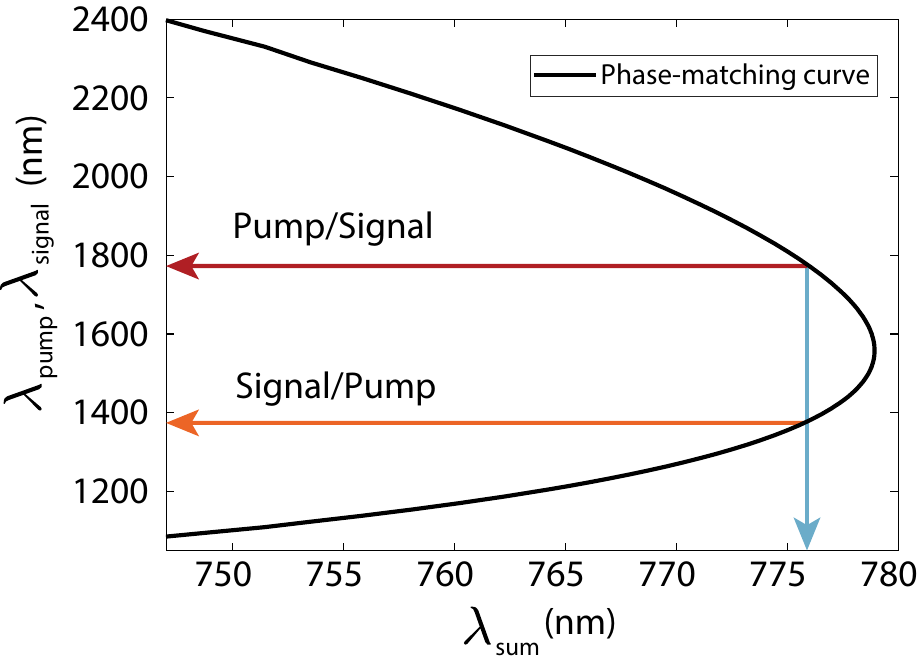}
		\caption{Simulated QFC phase-matching wavelengths for the fabricated waveguide.}
		\label{fig::S1}
	\end{center}
\end{figure*}

Fig.~\ref{fig::S1} shows the simulated QFC phase-matching wavelengths for the fabricated waveguide. They are determined through the frequency- and phase-matching conditions:  
	\begin{equation}
	\frac{1}{\lambda_{\text{pump}}} + \frac{1}{\lambda_{\text{signal}}} = \frac{1}{\lambda_{\text{sum}}},
	\qquad
	\frac{n_{\text{pump}}}{\lambda_{\text{pump}}} + \frac{n_{\text{signal}}}{\lambda_{\text{signal}}} = \frac{n_{\text{sum}}}{\lambda_{\text{sum}}},
	\end{equation}
where $n_k$ denotes the effective index of the corresponding mode and wavelength. In the simulation, we also included the material dispersion. Fig.~\ref{fig::S1} does not show the full phase-matching wavelength range, but the range of interest.

\section{Measurement details and data analysis}

\subsection{Phase-matching tuning and conversion efficiency measurement}

%\subsubsection{Measurement details}
We measure the sum frequency generation (SFG) and difference frequency generation (DFG) in the same propagation direction of the waveguide. Due to the different waveguide losses in the 1550-nm and 780-nm bands, as well as phase mismatch, the spectra of SFG and DFG need to be optimized separately. Consider a waveguide with two segments of the same length: the first segment is completely phase-mismatched, while the second is perfectly phase-matched. Because of the large loss in the 780-nm band, the DFG at the end of the waveguide is weaker than the SFG, because the 780-nm signal is lost without conversion in the first segment. If the two segments are switched, the SFG will be weaker than DFG.  
Since a perfect phase-matching condition cannot be achieved across the entire waveguide, which can be regarded as a combination of many phase-matched and mismatched segments, the spectra and efficiencies of SFG and DFG are inevitably different. 

\begin{figure*}[!htb]
	\begin{center}
		\includegraphics[width=1\columnwidth]{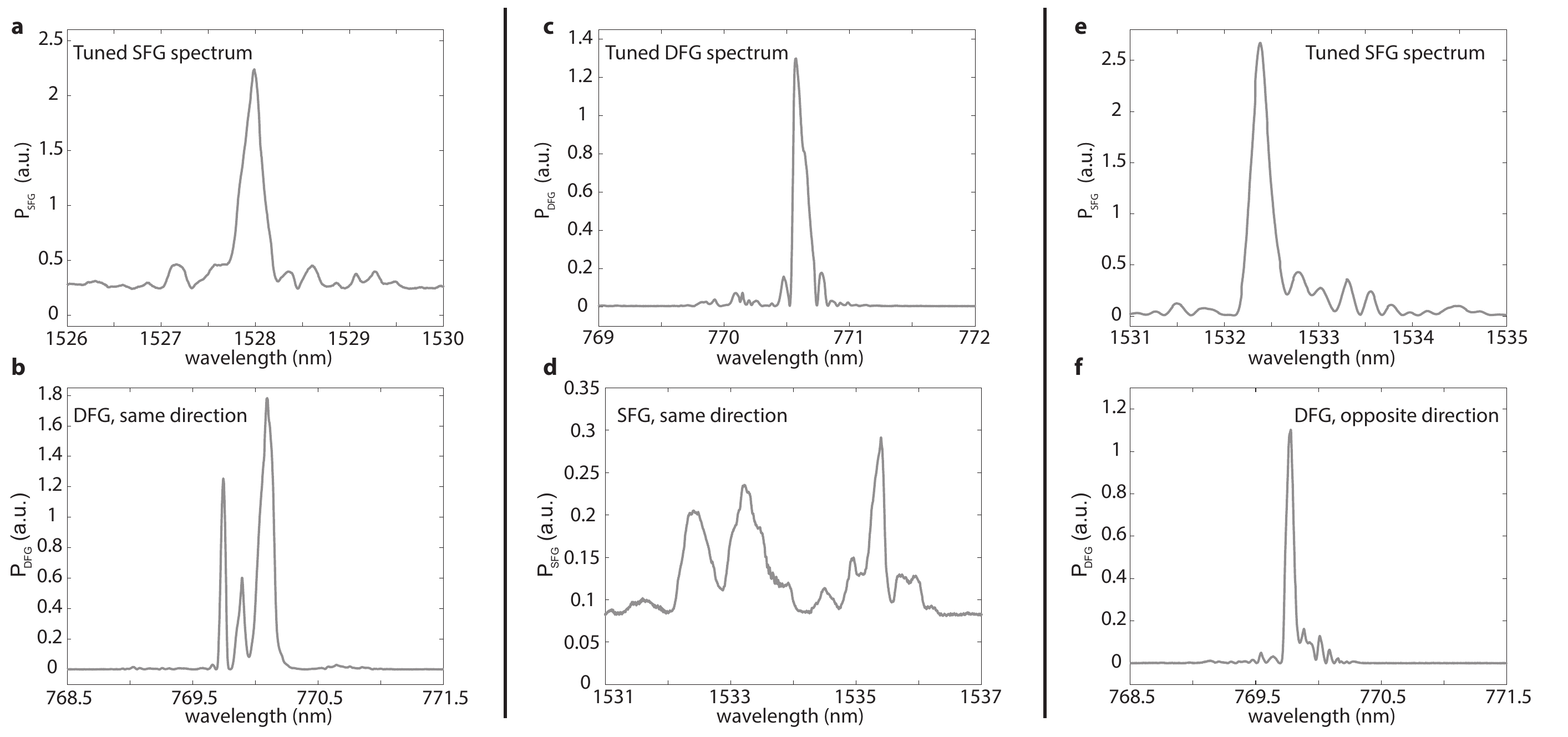}
		\caption{\textbf{Optimization of SFG and DFG.} 
			\textbf{a}. Tuned SFG spectrum. \textbf{b}. Same-direction DFG spectrum under the same condition. \textbf{c}. Tuned DFG spectrum. \textbf{d}. Same-direction SFG spectrum under the same condition. \textbf{e}. Tuned SFG spectrum. \textbf{f}. Opposite-direction DFG spectrum under the same condition. }
		\label{fig::S2}
	\end{center}
\end{figure*}

We use the nano-heaters to tune the phase-matching condition of the waveguide. Each nano-heater is 0.4-mm long, 400-nm wide, and 25-nm thick, which is individually controlled by a channel of a DC voltage source (NI 9264, 16-channel analog output module). Because of the large resistance of the nano-heater and the proximity to the waveguide, voltage of only up to a few volts is needed for each heater for the tuning.  
For SFG tuning, we start with monitoring the SFG signal in the 780-nm band and tune the first segment near the waveguide input, isolating the SFG response of the first segment from the rest of the spectrum. Then, the nano-heater of the second segment is used to align its SFG spectrum with that of the first segment to maximize the combined SFG peak intensity. This procedure is repeated sequentially for the remaining segments until every segment of the waveguide has been tuned at least once. Next, we switch the detector to monitor the residual 1550-nm signal in order to tune the dip of the 1550-nm band transmission, making it as deep as possible. After both both tunings, the phase matching for SFG is considered optimized, and we then vary the pump to measure the SFG conversion efficiency and the residual signal. 
For DFG tuning, we begin by monitoring the DFG signal in the 1550-nm band and start tuning from the first segment near the waveguide output side, aligning the DFG spectrum of each segment one by one.  

Because of the different loss of 1550-nm and 780-nm modes and the existence of phase mismatch, SFG and DFG in the same direction needs to be tuned separately. When one is optimized, the other one is not. Figs.~\ref{fig::S2}a and b show the optimized SFG and same-direction DFG under the same nano-heater condition, respectively. Figs.~\ref{fig::S2}c and d shows the optimized DFG and same-direction SFG under the same  nano-heater  condition, respectively.  It turns out the opposite-direction SFG and DFG can be simultaneously optimized as shown in Figs.~\ref{fig::S2}e and f.

\subsection{780-nm-band waveguide loss}

We also found, by isolating the SHG signal of each waveguide segment heated individually, the 780-nm-band waveguide loss can be measured precisely without calibration of the fiber couping efficiency, which can be difficult because of the large 780-nm transmission fringes.  Assuming each segment has similar phase matching quality, then the decay of the SHG signal of each segment reveals the loss of the 780-nm-band light (the loss of the 1550-nm-band pump is negligible). The loss of the 780-nm-band TM$_{00}$ mode is found to be 20 dB/cm in this way, which is consistent with microring's $Q_{780}\sim 10^5$ and simulated group index $n_g=5.7$ (i.e., $\beta=\frac{n_g\omega}{cQ}$). 

\begin{figure*}[!htb]
	\begin{center}
		\includegraphics[width=1\columnwidth]{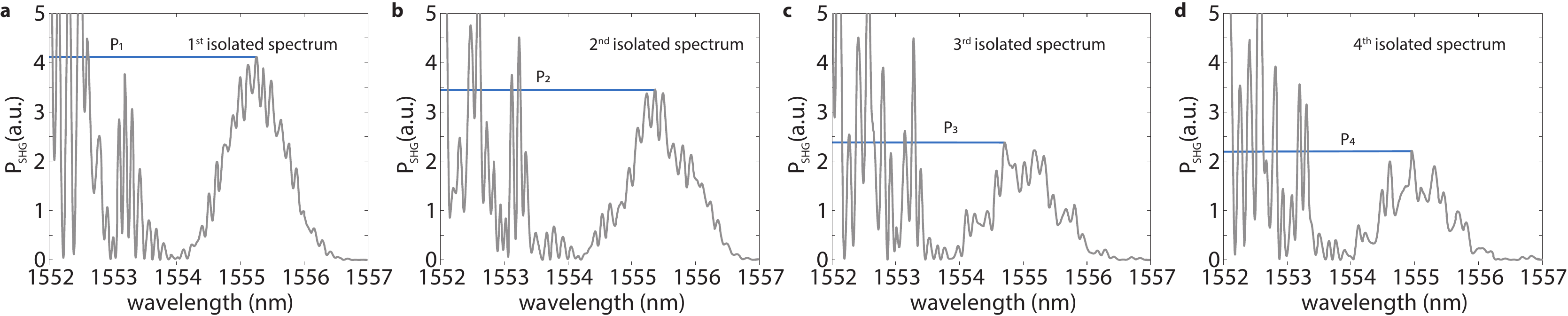}
		\caption{Isolated SHG spectrum of individual waveguide segment. }
		\label{fig::S3}
	\end{center}
\end{figure*}

Fig.~\ref{fig::S3} shows the isolated SHG spectrum of a few segments. The loss is calculated by  
\be\label{780loss}
\mathrm{Loss}=[10\mathrm{log}\frac{\sum_{i=1}^{i=N}(P_i/P_{i+1})}{N}]/L,
\ee
where $P_i$ is the peak SHG power of $i$-th segment and $N$ is the number of segments in waveguide length $L$. 

\subsection{Up-conversion and two-photon interference}
\subsubsection{Measurement details}
For the preparation of the entanglement source (ES), we used a continuous-wave (CW) tunable diode laser, locked to 770.95~nm by a wavelength meter, as the pump for spontaneous parametric down-conversion (SPDC). The laser output was amplified by a tapered amplifier to 36~mW. The idler and signal photons, were selected using an fiber Fabry-Perot (FFP) filter and two tunable filters (TF). The bandwidth and free spectral range of the FFP filter are 2.3 GHz and 1147 GHz, respectively. We used two nearby resonances of the FFP filter to select the entangled photon pair. A 20 nm bandwidth coarse wavelength division-multiplexer (CWDM) was then used to separate the FFP filter locking laser, with each channel containing at least two resonances of the FFP filter. Therefore, two narrowband TF were employed to isolate the signal and idler photons at wavelengths of 1537 nm and 1547 nm, respectively.
	
After the ES, the idler photon passes through the glass Mach--Zehnder interferometer (MZI) and is detected by superconducting nanowire single-photon detector (SNSPD, Quantum Opus). The signal photon is combined with a pump at 1567.04 nm and sent for frequency conversion in the InGaP waveguide. The converted signal photon in the 780-nm band then passes through the free-space interferometer (FSI) and is detected by a single-photon avalanche diode (SPAD). To measure the coincidence counts between the idler and signal photons, we use a time-correlated single photon counting module (Swabian) to calculate the correlation. Both the 780-nm pump laser and the 1550 nm locking	laser are stabilized using a wavelength meter. The 1550 nm locking laser is also used to stabilize the FFP filter and the FSI. The bandwidth of the quantum frequency conversion (QFC) is estimated based on the sum-frequency generation bandwidth of the devices used in this experiment, which is approximately 0.2 nm. Therefore, the 1550 nm pump laser was not stabilized using a wavelength meter.

The up-converted signal photon and the original idler photon pass through a free-space Michaelson interferometer and an integrated glass MZI, respectively, both with a time delay of $\tau=1$ ns, which is longer than the coherence time of the photons determined by the filter bandwidth. The photon pair can travel through either the short $(s)$ or long $(l)$ path of the unbalanced interferometers, forming an entangled state $\left|\psi\right\rangle=\frac{1}{2}(\left|s\right\rangle_1 \left|s\right\rangle_2 +e^{i(\varphi_1+\varphi_2)} \left|l\right\rangle_1 \left|l\right\rangle_2)$, where $\varphi_{1,2}$ is the path phase difference of the two interferometers. The entangled state can be post-selected using the time-resolved coincidence detection. 
The central peak corresponds to the coincidence probability of $\frac{1}{4}|1+e^{i(\varphi_1+\varphi_2)}|^2$. 

At the beginning of the experiment, we optimized the visibilities of interferometers using their corresponding continuous-wave lasers. The visibility of the glass MZI is influenced by the polarization of the input light and is limited by fabrication imperfections. The visibility of the FSI is polarization-independent and is primarily determined by the alignment of its internal components. The measured visibilities of the glass MZI are $V_{\text{1550,idler}} = 94.4\%$ and $V_{\text{1550,signal}} = 98.32\%$, while the visibility of the FSI is $V_{\text{FSI}} = 89\%$.

The two-photon visibilities are $V_{\text{1550-1550}} = 82.9\%$ and $V_{\text{1550-780}} = 80.0\%$. The measured visibilities are limited by the imperfect 50/50 splitting ratios of the beam splitters in the MZI and FSI. Taking these imperfections into account, the minimum theoretical visibilities are calculated to be $V_{\text{th,1550-1550}} = 86.59\%$ and $V_{\text{th,1550-780}} = 72.44\%$ (see the next section). The measured $V_{\text{1550-1550}}$ is slightly lower than the theoretical value, which is likely due to added noise and system losses. Interestingly, the measured $V_{\text{1550-780}}$ exceeds the minimum theoretical value. We attribute this to the spectral filtering effect introduced by the QFC process, which is likely to enhance visibility. Similar improvements in the quantum properties of light during QFC have also been reported in previous studies \cite{zaske2012visible,ates2012two}

\subsubsection{Modeling of two-photon fringe visibility}

Consider two unbalanced MZI, each consisting of two beam splitters, with transmission and reflection coefficients denoted by $T_{i1(i2,s1,s2)}$ and $R_{i1(i2,s1,s2)}$, respectively. These coefficients satisfy the relation $T_{jk}^2 + R_{jk}^2 = 1$. Here, $i$ and $s$ represent the idler and signal photons, respectively. For a single-photon input state, the output state of the MZI is given by
\begin{equation}\label{eq:SPInput}
	\begin{split}
		\ket{1}_i &\to T_{i1}T_{i2}\ket{s}_i + R_{i1}R_{i2}e^{i\varphi_s}\ket{l}_i \\
		\ket{1}_s &\to T_{s1}T_{s2}\ket{s}_s + R_{s1}R_{s2}e^{i\varphi_i}\ket{l}_s ,
	\end{split}
\end{equation}
where $\ket{s}$ and $\ket{l}$ denote the state in the short and long path of the interferometer, and $\varphi_{s,i}$ is the phase difference of the two paths. The visibilities of the single-photon interference thus are 

\begin{equation}\label{eq:SPVisibility}
	\begin{split}
		V_{i1}&=\frac{2T_{i1}T_{i2}R_{i1}R_{i2}}{T_{i1}^{2}T_{i2}^{2}+R_{i1}^{2}R_{i2}^{2}} \\
		V_{s1}&=\frac{2T_{s_{1}}T_{s_{2}}R_{s_{1}}R_{s_{2}}}{T_{s_{1}}^{2}T_{s_{2}}^{2}+R_{s_{2}}^{2}R_{s_{2}}^{2}} ,
	\end{split}
\end{equation}
The single photon interference visibility can be measured via CW lasers.

After passing the interferometer, the time-resolved entangled signal-idler photon pair state is given by
\begin{equation}\label{eq:TwoPhotonState}
	\ket{1}_i\ket{1}_s \to T_{i1}T_{i2}T_{s1}T_{s2}\ket{s}_i\ket{s}_s + R_{i1}R_{i2}R_{s1}R_{s2}e^{2i\varphi}\ket{l}_i\ket{l}_s,
\end{equation}
where $\varphi$ is the interferometer phase difference corresponding to half of the SPDC pump frequency. The visibility of the two-photon interference is given by
\begin{equation}\label{eq:TwoPhotonVisibility}
	V_{2}=\frac{2T_{i1}T_{i2}T_{s1}T_{s2}R_{i1}R_{i2}R_{s1}R_{s2}}{T_{i1}^{2}T_{i2}^{2}T_{s1}^{2}T_{s2}^{2}+R_{i1}^{2}R_{i2}^{2}R_{s1}^{2}R_{s2}^{2}},
\end{equation}
Using Eqs. \ref{eq:SPVisibility} and \ref{eq:TwoPhotonVisibility}, the single and two-photon visibility are related by
\begin{equation}\label{eq:VisibilityRelation}
	V_{2}=\frac{2}{\frac{2}{V_{i1}V_{s1}}-\frac{T_{i1}^2T_{i2}^2R_{s1}^2R_{s2}^2+T_{s1}^2T_{s2}^2R_{i1}^2R_{i2}^2}{T_{i1}T_{i2}R_{s1}R_{s2}T_{s1}T_{s2}R_{i1}R_{i2}}}\geq \frac{2}{\frac{2}{V_{i1}V_{s1}}-2}.
\end{equation}
Eq. \ref{eq:VisibilityRelation} not only applies to two MZIs but also applies to one MZI and one FSI case. Based on the measured single-photon visibilities, $V_{\text{1550,idler}} = 94.4\%$, $V_{\text{1550,signal}} = 98.32\%$, and $V_{\text{FSI}} = 89\%$, the minimum theoretical two-photon visibilities are:
\begin{equation}\label{eq:TheoryVisibility}
	\begin{split}
		V_{th,1550-1550}&=\frac{2}{\frac{2}{V_{1550,idler}V_{1550,signal}}-2}=86.59\% \\
		V_{th,1550-780} &=\frac{2}{\frac{2}{V_{1550,idler}V_{FSI}}-2}=72.44\%.
	\end{split}
\end{equation}

\subsection{Down-conversion and quantum state tomography}

The 1550\,nm pulses were generated by intensity-modulating a continuous‐wave laser, wavelength-locked at 1551.91 nm. The resulting pulses have a full‐width at half maximum (FWHM) of 434\,ps and a repetition rate of 250\,MHz. These pulses are up‐converted to 780-nm-band photon via second harmonic generation (SHG) using a PPLN waveguide.  After SHG, the 780\,nm pulses pass through an unbalanced FSI. The FSI is actively stabilized using a feedback loop to maintain the desired phase delay, yielding the final time‐bin state. A second 1550\,nm pump laser locked at 1564.04 nm is combined with the 780-nm-band pulse through WDM and sent into the wavelength-conversion device. Before coupling, the amplified spontaneous emission (ASE) noise is suppressed by a WDM filter in the 1570\,nm band. At the device output, residual pump light and 780-nm-band signal light are removed by dichroic filters, and noise photons are further rejected by a TF with a 473\,GHz passband, centered on the converted wavelength.

We perform projection measurements onto the six time‐bin basis states
\[
|0\rangle\langle0|,\;|1\rangle\langle1|,\;|+\rangle\langle+|,\;|-\rangle\langle-|\,,\;|L\rangle\langle L|,\;|R\rangle\langle R|\;,
\]
denoted by counts $N_0$, $N_1$, $N_+$, $N_-$, $N_L$, and $N_R$, respectively (here $|0\rangle\equiv|e\rangle$ and $|1\rangle\equiv|l\rangle$). The Stokes parameters for a time‐bin qubit are then
\begin{equation}\begin{gathered}
		S_0=N_0+N_1, \\
		S_{1}=N_{+}-N_{-}, \\
		S_{2}=N_{L}-N_{R}, \\
		S_3=N_0-N_1.
\end{gathered}\end{equation}
The density matrix is reconstructed as
\begin{equation}\rho=\frac{1}{2}(\sigma_0+\frac{S_1}{S_0}\sigma_1+\frac{S_2}{S_0}\sigma_2+\frac{S_3}{S_0}\sigma_3),\end{equation}
where $\sigma_i (i=0, 1, 2, 3)$ is Pauli matrices.

In our experiment, each projection outcome is obtained from photon counts in the three temporal modes at the output of the MZI: the early (\(e\)), late (\(l\)) and late-late (\(ll\)) time bins. We denote by \(n_{\varphi,j}\) the SNSPD counts in bin \(j\in\{e,l,ll\}\) when the MZI phase is set to \(\varphi\).  The total counts for the six projectors are then
\begin{equation}\begin{gathered}
		N_{0}=n_{0,e}+n_{\pi,e}, \\
		N_{1}=n_{0,ll}+n_{\pi,ll}, \\
		N_-=n_{0,l}, \\
		N_{+}=n_{\pi,l}, \\
		N_{L}=n_{\frac{\pi}{2},l}, \\
		N_{R}=n_{\frac{3\pi}{2},l},
\end{gathered}\end{equation}

Here \(N_{0}\) and \(N_{1}\) correspond to the \(\lvert0\rangle\langle0\rvert\) and \(\lvert1\rangle\langle1\rvert\) projections (measured in the early and late bins, respectively), while \(N_{\pm}\) and \(N_{L/R}\) are the superposition‐basis counts extracted from the central bin at phases \(0,\pi,\tfrac{\pi}{2}\) and \(\tfrac{3\pi}{2}\), respectively.

The directly measured projected counts can result in an unphysical density matrix due to measurement imperfections. A maximum likelihood estimation algorithm \cite{altepeter2005photonic} constructs the physical density matrix of each measured state. Using the input state $\vert \psi \rangle$ the fidelity can then be calculated as $\mathcal{F} =\langle \psi \vert \rho \vert \psi \rangle$.

\section{Noise measurement and analysis}

\subsubsection{Measurement of on-chip noise}
To measure the on-chip noise, only pump is injected into the device. To characterize the backward scattering noise generated in the waveguide, the reflected light from the device is filtered by a CWDM and a 4-nm-bandwidth TF to separate the residual pump, and subsequently detected using an SNSPD. The pump is set to a wavelength of 1543~nm, and its power is varied. Measurements are taken from both the 1530-nm and 1570-nm ports of the CWDM, where the count rates are found to be similar. The backward on-chip noise is obtained through a differential method by subtracting the noise for fiber coupled and decoupled with the device, to remove noise that originates outside the device, such as in the fiber. To measure the forward noise generated in the waveguide, similar setup is used. However, the differential method is changed to subtraction of the noise of a long and a short waveguide.

\subsubsection{Spontaneous four-wave mixing noise}
\begin{figure*}[!htb]
	\begin{center}
		\includegraphics[width=1\columnwidth]{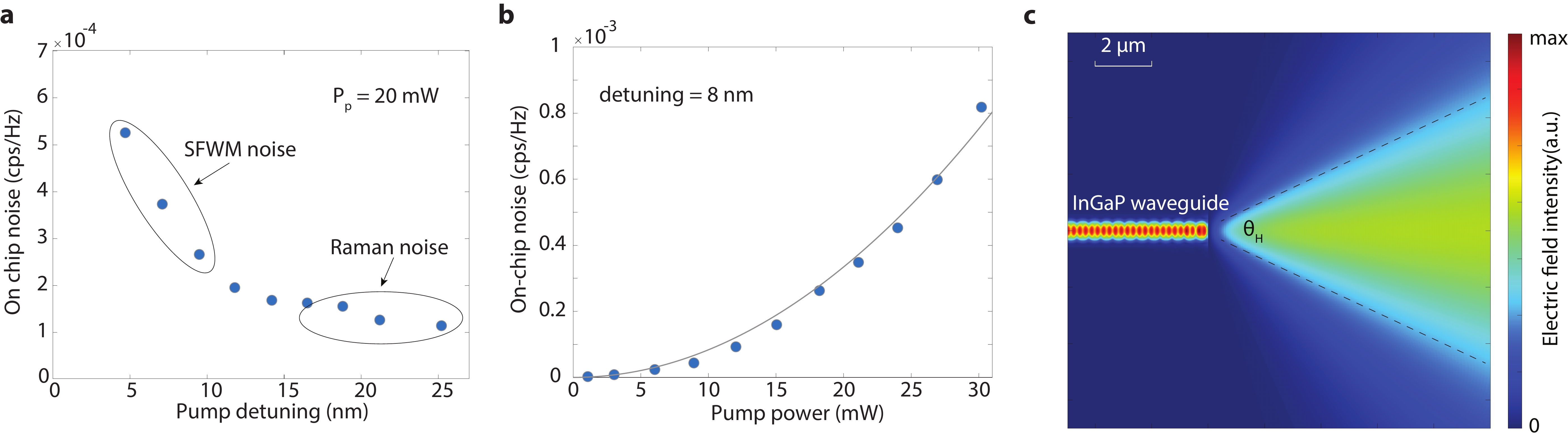}
		\caption{\textbf{On-chip noise}. \textbf{a}. Forward on-chip noise versus pump detuning with 20 mW pump power. \textbf{b}. Forward on-chip noise versus pump pump power for pump detuning of 8 nm.  
			\textbf{c}. Simulated beam emission from the InGaP waveguide.  1550-nm TE$_{00}$ mode was excited. Electrical field intensity outside the waveguide is rescaled.}
		\label{fig::S4}
	\end{center}
\end{figure*}

The measured forward noise versus pump detuning, i.e., wavelength detuning from the pump, with 20 mW pump power is shown in Fig. \ref{fig::S4}a. Fig. \ref{fig::S4}b shows the forward noise versus pump power for detuning of 8 nm, which reveals a quadratic dependence. Such quadratic noise is due to the degenerate spontaneous four-wave mixing (SFWM) of the pump. This is consistent with the simulated bandwidth of the degenerate SFWM of 30 nm for the 6-mm waveguide and the estimated noise rate $(\eta P_pL)^2$ (cps/Hz) \cite{helt2012does}, where $\eta\approx 200$/W/m is the nonlinear coefficient. The quadratic noise cannot be due to the cascaded $\chi^{(2)}$ processes of pump SHG $\rightarrow$ SPDC, because the pump detuning here is much larger than the phase-matching bandwidth of such processes, which is a few tens of GHz.

\subsubsection{Modeling spontaneous Raman scattering noise in the waveguide}
When the pump detuning is larger, e.g., 20 nm, the noise becomes linearly dependent on the pump power, as shown in Fig. 2f. Because the noise is linear in pump power and the pump frequency is below the bandgap, it could be attributed to spontaneous Raman scattering or defect-induced fluorescence. Though the two noise sources can be distinguished based on the fact that spontaneous Raman scattered light is polarized while fluorescence is not, our InGaP waveguide only support TE-modes in the 1550-nm band and consequently we cannot distinguish the two types of noises based on polarization---if both of them exist. However, since InGaP is an undoped crystalline material, unlike amorphous materials, defect-induced florescence is expected to be low.  Below we model the spontaneous Raman scattering induced noise.

The spontaneous Raman scattering photon flux in a waveguide can be calculated by \cite{kullander2015crosssection}
\begin{equation}\label{RS}
	F_{\mathrm{RS}} =\beta \Omega \frac{d\sigma_{\mathrm{eff}}}{d\Omega}\frac{M }{\Delta \nu}LF_{p},
\end{equation}
where $\sigma_{\mathrm{eff}}$ is the effective Raman cross section in the waveguide, $\Delta \nu$ is the bandwidth of the Raman peak, $\Omega$ is the solid angle of Raman scattering in the waveguide, $L$ is the waveguide length, $M$ is the molecular density, and $F_{p}$ is pump photon flux. We further introduced a factor $\beta$ to account for the ratio between the detuned Raman background and the Raman peak. The effective Raman cross section of the waveguide, $\sigma_{\mathrm{eff}} $, can be related to that of a bulk material, $\sigma$, via \cite{holmstrom2016effcrosssection}:
\begin{equation}
	\sigma_{\mathrm{eff}} = \frac{n_{\mathrm{RS}}^2 \lambda_s\lambda_p}{8\pi} 
	\frac{n_g \iint_{\mathrm{WG}} |E(x,y)|^4 \,\mathrm{d}x \,\mathrm{d}y}
	{\left(\iint_{\infty} n^2 |E(x,y)|^2 \,\mathrm{d}x \,\mathrm{d}y \right)^2} \sigma,
\end{equation}
where $n_{\mathrm{RS}}$ is the refractive index of the Raman medium,  $n_g$ is group index of the waveguide, $\lambda_p$ is the pump wavelength, $\lambda_s$ is the wavelength of the Raman peak, $E(x,y)$ is the transverse electric field distribution of the waveguide mode, and the integral in the denominator is performed for whole space with $n$ the refractive index of individual material. 

The spontaneous Raman scattering cross section can be related to the peak stimulated Raman gain $g$ via \cite{demos2011gaincrosssection}
\begin{equation}\label{g}
	g = \frac{8 \pi c^2 M}{\hbar \omega_s^3 n_{\mathrm{eff}}^2 (N_0+1)\Delta\nu}  \frac{d\sigma_{\mathrm{eff}}}{d\Omega},
\end{equation}
where $\omega_s$ is the angular frequency of the Raman peak, $n_{\mathrm{eff}}$ is the effective index of the waveguide, $N_0=e^{-\hbar\omega_{\mathrm{ph}}/k_BT} $ is the Bolzman factor (where $\omega_{\mathrm{ph}}$ is the phonon frequency). 
Using Eqs. \ref{RS} and \ref{g}, we can re-write the spontaneous Raman scattering photon flux in terms of peak Raman gain as
\be
F_{\mathrm{RS}} =\frac{\hbar \omega_s^3 n_{\mathrm{eff}}^2 (N_0+1)}{8 \pi c^2}L\beta \Omega gF_{p},
\ee

There remains three variables to be determined to estimate $F_{\mathrm{RS}}$ of the InGaP waveguide: the peak Raman gain $g$, the solid angle $\Omega$, and the background-to-peak ratio $\beta$. 
\begin{itemize}
\item Peak Raman gain $g$: we do not find $g$ for InGaP waveguides in the literature. Instead, we use $g=1.23 \cross 10^{-9} \mathrm{m/W}$ of GaP waveguides \cite{saito2000GaPraman} as an approximation.

\item Solid angle $\Omega$: we performed 3D finite-difference time-domain simulation to calculate the radiation angle of the TE$_{00}$ mode of the InGaP waveguide. As shown in Fig. \ref{fig::S4}c, we obtained beam spread angle (in radiant) $\theta_H = 0.87$ (in plane) and $\theta_V = 1.00$ (out of plane). Applying Snell's law at the InGaP-air interface, the effective solid angle inside the waveguide is calculated as $\Omega \approx\frac{\pi}{4} \theta_H\theta_V/n^2= 0.0662~\mathrm{sr}$.

\item  Background-to-peak ratio $\beta$: Ref. \cite{yue2015Ramanratio}  measured Raman spectrum of InGaP, from which we inferred $\beta=0.127$. 

\end{itemize}
Using these parameters, the Raman noise is estimated to be $F_{\mr{RS}}=2 \times 10^{-3}$ cps/Hz.

Given the measured noise in the 1550-nm band, we estimate the up-converted noise in the 780-nm band using 
\begin{equation}
	F_{U}={\int_0^L \frac{F_{\mathrm{RS}}}{L}\eta(z)\mathrm{d}z},
\end{equation}
where $\eta$ is the conversion efficiency given by Eq. \ref{CE} and we have used the fact that $F_{\mathrm{RS}}$ is proportional to $L$. Using the measured $F_{\mathrm{RS}}=10^{-4}$ cps/Hz, we find the up-converted Raman noise would be $F_{U}=2\times 10^{-5}$ cps/Hz.

\section{Self-calibrated SFG efficiency}

QFC conversion efficiency is related to the normalized SFG efficiency in the weak pump limit, $\eta_{\mathrm{SFG}}$, given by Eq.~\ref{CE}. By fitting the measured QFC conversion efficiency versus pump power, we found $\eta_{\mathrm{SFG}}=700~ \text{W}^{-1}\text{cm}^{-2}$ and $1750~ \text{W}^{-1}\text{cm}^{-2}$ for the 6-mm and 2.5-mm waveguides, respectively. The $\eta_\text{SFG}$ for the 6-mm-long waveguide is less than that of the 2.5-mm-long waveguide, and both are significantly lower than the simulated efficiency of $5200~ \text{W}^{-1}\text{cm}^{-2}$, indicating there is still substantial phase mismatch in the waveguides.

To quantify the phase mismatch of a waveguide, one can calculate the self-calibrated efficiency ratio defined as \cite{chen2024adapted,nash1970effect},
\begin{equation}\label{eq:RDefinition}
	R\equiv\frac{\eta}{\eta_0}=\frac{\eta\alpha}{AL},
\end{equation}
where $\eta$ and $\eta_0$ denote the peak efficiency of the measured and ideal frequency conversion spectrum, e.g., SFG or SHG, respectively, $A=\int \eta(\lambda)d\lambda$ and $\alpha=2\pi(\text{d}\Delta k/ \text{d}\lambda)^{-1}$ is the simulated dispersion factor with $\Delta k$ the wavevector mismatch between the modes. It should be noted that $\lambda$ in the definition of $A$ and $\alpha$ corresponds to the same mode. Because $R$ is related to the ratio of $\eta$ and $A$, it is free of calibration of the fiber coupling efficiency as well as the waveguide loss \cite{nash1970effect}. In addition, the bandwidth of the ideal frequency conversion spectrum, denoted as $B_0$, follows the relation $B_0=\frac{5.57}{L}(\frac{\text{d}\Delta k}{\text{d}\Delta \lambda})^{-1}$ \cite{helmfrid1993influence}. The measured bandwidth $B$ is generally larger than $B_0$ due to the phase mismatch in the waveguide. 

\begin{figure*}[!htb]
	\begin{center}
		\includegraphics[width=1\columnwidth]{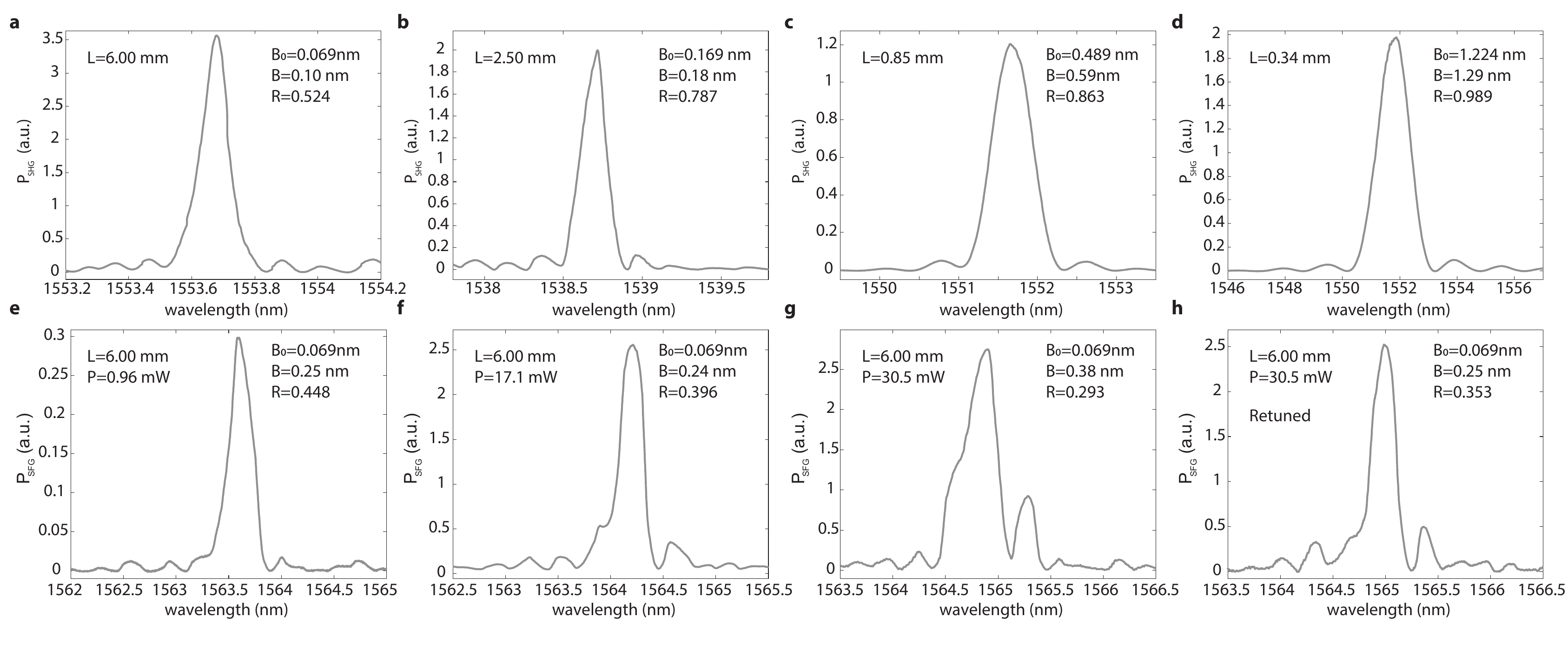}
		\caption{\textbf{Measured SHG and SFG spectrum.} 
			\textbf{a-d}. SHG spectra for waveguides of different lengths. Spectra for the 6-mm and 2.5-mm waveguides are after tuning. \textbf{e-g}. SFG spectra for a 6-mm waveguide under varying pump power. \textbf{h}. Re-tuned SFG spectrum for the 6-mm waveguide at high power. }
		\label{fig::S5}
	\end{center}
\end{figure*}

Figs. \ref{fig::S5}a-d present the measured SHG spectrum of waveguides with different lengths. The 6-mm and 2.5-mm waveguides are tuned with the nano-heaters while  the 0.85-mm and 0.34-mm waveguides are not tuned. As the waveguide length increases, the value of $R$ decreases, even with tuning for the longer waveguides. This trend is also reflected in the reduced bandwidth ratio $B/B_0$. This indicates that there is phase mismatch within individual waveguide segment that cannot be eliminated by the nanoheater tuning. In addition, thermo-optic effect induced by high pump power further degrade the phase-matching condition. Figs. \ref{fig::S5}e-g show SFG spectra of the 6-mm waveguide with increasing pump power, that is initially tuned at a low pump power. The value of $R$ decreases as the pump power increases. Even after re-tuning at high power, as shown in Fig. \ref{fig::S5}h, the value of $R$ remains less than in the low-power case. 

If we use $R=0.37$ and $R=0.83$ for the 6-mm and 2.5-mm waveguides, respectively, averaged for various pump powers used in the frequency conversion, the corrected values of the normalized SFG efficiencies are $\eta_{0} = 700/0.37 = 1892~\text{W}^{-1}\text{cm}^{-2}$ and $\eta_{0} = 1750/0.83 = 2108~\text{W}^{-1}\text{cm}^{-2}$ for the 6-mm and 2.5-mm waveguides, respectively, which are comparable to the simulated value. However, the cause of the remaining discrepancy is unclear yet. 

\vspace{2mm}